\documentstyle[aps,pra,eqsecnum]{revtex}
\begin{document}
\bibliographystyle{unsrt}
\title {Propagation of a relativistic particle in terms  of the unitary irreducible representations of the Lorentz group}  
\date{\today}
\author{Rudolf A. Frick  \thanks{Email: rf@thp.uni-koeln.de}\\{ Institut f\"ur Theoretische Physik, Universit\"at K\"oln, D-50923 Cologne, Germany} \\}

\maketitle
\begin{abstract}
In  a generalized Heisenberg/Schr\"odinger picture we use an invariant space-time transformation  to describe the  motion of a relativistic particle. We discuss the relation with the relativistic  mechanics and find that  the propagation  of the particle  may be defined as  space-time transition between states   with equal eigenvalues of the first and second Casimir operators of the Lorentz algebra.   In addition we use a vector  on the  light-cone. A massive relativistic particle with spin 0 is considered. We also consider   the nonrelativistic limit.
\end{abstract}{\it PACS numbers: 03.65. Pm,  03.65. -w}
\section{Introduction}  In this  paper  we  present a new  mathematical formalism    for describing  the motion of a relativistic particle   which is based  on the principal series of the unitary  irreducible representations of the Lorentz group and  a  generalized Heisenberg/Schr\"odinger picture.    The principal series of the representations of the Lorentz group has already been used by many authors in the theory of elementary particles and relativistic nuclear  physics (e.g., Refs. \cite{Joos,Ruhl,Kad,Barut,Ska,Shapiro,Kar}). In our previous   papers  \cite{Frick1,Frick2,Frick3} it has been shown that  these representations may  be used in a generalized Heisenberg/Schr\"odinger picture in which either the analogue of Heisenberg states or the analogue of Schr\"odinger operators are independent of both time and space coordinates t, ${\bf x}$. For these states there must be space-time independent expansion. If   at first we use the momentum representation in  the  expansion  of the Lorentz group,  then the states and operators of the Poincar\'e algebra can be constructed in another space-time independent representation.

 In  \cite{Frick1} the transition from the Heisenberg   to the Schr\"odinger picture in  quantum mechanics $S(t)=\exp(-itH)$ was generalized  to the  relativistic invariant transformation (we choose here a system of units such that ${\hbar}=1,{\,}c=1)$
\begin{equation}
\label{1.1}
 S(x):=\exp[-i(tH-{\bf x}\cdot{\bf P})],
\end{equation} 
 where   ${H}$ and ${\bf P}$ are the  Hamilton and momentum operators of the particle in the generalized Schr\"odinger picture.   Through this transformation  the plane waves $\sim\exp[-ixp]$ appear in   different representation and  cannot be used   in their original sense as the stationary   states of a particle. There is no ${\bf x}$ representation. In this approach one must find a new method  for describing  the  motion  of the particle.

In the present paper we will  show that using the transformation $S(x)$ makes it possible to  describe  the motion of a massive relativistic particle in  terms of the matrix elements  of the Lorentz group. First we  introduce in the   relativistic mechanics  the analogue of the operators  of the Lorentz algebra in the  generalized Heisenberg picture and  obtain equations which in the invariant form express    the  motion of a particle.  Then we use  these equations  and  the  unitary irreducible representations of the Lorentz group to  determine  the transiton amplitudes for the free relativistic particle. In particular we consider a massive particle with spin 0.  In  the nonrelativistic limit we use the   expansion of the Galileo group.
\section{ Lorentz algebra in  the generalized Heisenberg picture. Analogue in  relativistic mechanics}
In the generalized Heisenberg/Schr\"odinger picture the analogue of Schr\"odinger operators of a particle are defined as space-time  independent operators in different representations. In the momentum representation (${\bf p}$ = momentum, m = mass, $p_0:=\sqrt{m^2+{\bf p}^2}$,  s = spin ) the   boost and rotation generators of the Lorentz group
\begin{equation}
\label{2.1}
{\bf N}:=ip_0{\nabla}_{\bf p}-\frac{{\bf s}\times{\bf p}}{p_0+m},\quad{\bf J}=-i{\bf p}\times{\bf \nabla}_{\bf p}+{\bf s}:={\bf L}({\bf p})+{\bf s}.
\end{equation}
can be viewed as such operators.  
Using  the transformation $S(x)$ we obtain  the  operators of the Lorentz algebra in the  generalized Heisenberg picture
\begin{equation}
\label{2.2}
{\bf N}(x)=S^{-1}(x){\bf N}S(x)={\bf N}+t{\bf P}-{\bf x}{H},
\end{equation}
\begin{equation}
\label{2.3}
{\bf J}(x)=S^{-1}(x){\bf J}S(x)={\bf J}-{\bf x}\times{\bf P}.
\end{equation}
Time and space coordinates equally occur in these  operators.  From this point of view one can see ${\bf N}(x),\,{\bf J}(x)$  as field operators which satisfy the  equations 
\begin{equation}
\label{2.4}
\frac{\partial{N_i(x)}}{\partial{t}}=P_i,\quad\frac{\partial{N_i(x)}}{\partial{x_j}}=-H{\delta}_{ij},\quad\frac{\partial{J_i(x)}}{\partial{x_j}}=-{\epsilon}_{ijk}P_k,
\end{equation}
 and the commutation rules of the Lorentz algebra
\begin{eqnarray}
\label{2.5}
\lbrack{N_i(x)},{N_j(x)}\rbrack=-\imath\epsilon_{ijk}{J_k}
(x),\quad\lbrack{N_i(x)},{J_j(x)}\rbrack=\imath\epsilon_{ijk}{N_k}(x),
\end{eqnarray}
\begin{eqnarray}
\label{2.6}
\lbrack{J_i(x)},{J_j(x)}\rbrack=\imath\epsilon_{ijk}{J_k}(x).
\end{eqnarray}
For the   Casimir operators we have
\begin{eqnarray}
\label{2.7}
C_1(x):={\bf N}^2(x)-{\bf J}^2(x),\quad{C_2}(x):={\bf N}(x)\cdot{\bf J}(x).
\end{eqnarray} 
We introduce the  field ${\bf N}(x),\,{\bf J}(x)$  in the relativistic mechanics and use the same symbols. In the problem which we discuss one must  find the property of the field  ${\bf N}(x),\,{\bf J}(x)$ and the invariant $C_1(x),\,{C_2}(x)$ along the  trajectory of a particle.

Let us   write  (\ref{2.2}) and (\ref{2.3}) in the relativistic mechanics in the form
\begin{equation}
\label{2.8}
{\bf N}(x)={\bf N}(t_0,{\bf x}_0)+(t-t_0){\bf P}-({\bf x}-{\bf x}_0){H},
\end{equation}
\begin{equation}
\label{2.9}
{\bf J}(x)={\bf J}(t_0,{\bf x}_0)-(({\bf x}-{\bf x}_0)-(t-t_0){\bf P}/H)\times{\bf P},
\end{equation}
where ${\bf x}_0$  are the position of the particle on the time ${t_0}$.
For the  trajectoty  $({\bf x}_t={\bf x}_0+(t-t_0){\bf P}/H)$ we have
\begin{equation}
\label{2.10}
{\bf N}+t{\bf P}-{\bf x}_t{H}={\bf N}+t_0{\bf P}-{\bf x}_0{H},\quad\quad{\bf J}-{\bf x}_t\times{\bf P}={\bf J}-{\bf x}_0\times{\bf P}.
\end{equation}
In these formulas  the  quantity ${\bf N},\,{\bf J}$ are  separated from  integrals  of the motion $t_0{\bf P}-{\bf x}_0{H},\\{\bf x}_0\times{\bf P}$ because   the operators ${\bf N},\,{\bf J}$  in   the generalized Heisenberg/Schr\"odinger picture  correspond to  the space-time independent quantity.  For two points of the  trajectoty  we obtain
\begin{equation}
\label{2.11} 
{\bf N}(t_1,{\bf x}_1)={\bf N}(t_2,{\bf x}_2),\quad{\bf J}(t_1,{\bf x}_1)={\bf J}(t_2,{\bf x}_2),
\end{equation}
and come to the conclusion that 
\begin{equation}
\label{2.12} 
C_1(t_1,{\bf x}_1)=C_1(t_2,{\bf x}_2),\quad{C_2}(t_1,{\bf x}_1)=C_2(t_2,{\bf x}_2).
\end{equation}
These equations  represent  the motion of a particle from the point $t_1,{\bf x}_1$ to the point $t_2,{\bf x}_2$ in the invariant form and may be used in the quantum version.

In connection with  Eqs.(\ref{2.10}) we make the following remarks.  The space-time parts in (\ref{2.2}) and (\ref{2.3}) have the structure of the four-tensor of angular momentum of a particle. If we assume  contradictorally to  the concept of the generalized Heisenberg/Schr\"odinger picture  that the operators ${\bf N}, {\bf J}$ in the form 
\begin{equation}
\label{2.13}
{\bf x_t}={\bf N}/H+t{\bf P}/H,\quad{\bf J}={\bf x_t}\times{\bf P}
\end{equation}
 correspond  in the relativistic mechanics to  integral of the motion then  we arrive  at the ${\bf x}$ representation and 
\begin{equation}
\label{2.14}
{\bf N}(t,{\bf x}_t)=0,\quad{\bf J}(t,{\bf x}_t)=0,\quad{C_1}(t,{\bf x}_t)=0,\quad{C_2}(t,{\bf x}_t)=0.
\end{equation}
 In this case the zero on the right-hand side in (\ref{2.14}) makes  the inverse transition from the relativistic mechanics to the  quantum version  impossible.
\section{Transition  amplitudes} 
In the quantum version  the equations (\ref{2.12})  correspond  to  the transition $S(x_2-x_1)$ of the particle   from one state to another state with equal eigenvalues of the operators 
\begin{equation}
\label{3.1}
 C_1={\bf N}^2-{\bf J}^2,\quad C_2={\bf N}\cdot{\bf J}.
\end{equation} 
 For the  principal series   the eigenvalues  of  the operators  $C_1$ and $C_2$  are   $1+\alpha^2-{\lambda}^2$ and  $\alpha\lambda$, $(0\leq\alpha<\infty,\quad\lambda=-s,...,s)$, respectively. The representations $(\alpha,\lambda)$ and $(-\alpha,-\lambda)$ are unitarily equivalent.
In the momentum representations for a massive particle with spin zero we use the eigenfunctions of the operators $C_1$
\begin{equation}
\label{3.2}
\xi^{(0)}({\bf p},{\alpha},{\bf n}):=\frac{1}{(2\pi)^{3/2}}[(pn)/m]^{-1+i\alpha},
\end{equation} 
here  $n:=({\bf n},{n_0})$ is a  vector on the  light-cone $({\bf n}^{2}-n^2_0=0)$.  
These functions  were  used  first  in  Ref.\cite{Shapiro} in the space-time independent expansions of   the Lorentz group $( n_0=1, {\bf n}:=(\sin{\theta}\cos{\varphi},\sin{\theta}\sin{\varphi},\cos{\theta}))$
\begin{equation}
\label{3.3}
\Psi^{(0)}({\bf p})=\int{\alpha}^2d{\alpha}\,d{\omega}_{\bf n}\,\Psi^{(0)}({\alpha},{\bf n})\,\xi^{(0)}({\bf p},{\alpha},{\bf n}),
\end{equation}
\begin{equation}
\label{3.4}
\Psi^{(0)}({\alpha},{\bf n})=\int\frac{d{\bf p}}{p_0}\,\Psi^{(0)}({\bf p})\,\xi^{\ast(0)}({\bf p},{\alpha},{\bf n})
\end{equation}
where  $\Psi^{(0)}({\bf p})$ and  $\Psi^{(0)}({\alpha},{\bf n})$
are the state functions of the particle with spin zero in \({\bf p}\) and in the \({\alpha},{\bf n}\) representation. The  Hamilton operator $H^{(0)}(\alpha,{\bf n})$ and momentum operators ${\bf P}^{(0)}({\alpha},{\bf
  n})$ were constructed in Ref.\cite{Kad}. The operators ${\bf N}$ in the ${\alpha},{\bf n}$ representation have the form ( Ref.\cite{Frick1,Frick2,Frick3})
\begin{equation}
\label{3.5}
{\bf N}:=\alpha{\bf n}+({\bf n}\times{\bf L}-{\bf L}\times{\bf n})/2,\quad{\bf L}:={\bf L}({\bf n}).          
\end{equation}
For the particle with spin s
\begin{equation} 
\label{3.6}
{\bf N}:={\alpha}{\bf n}+({\bf n}\times{\bf J}-{\bf J}\times{\bf n})/2\quad{\bf J}:={\bf L}({\bf n})+{\bf s},
\end{equation}
\begin{equation}
\label{3.7}
 C_1=1+\alpha^2-({\bf s}\cdot{\bf n})^2,\quad{C}_2=\alpha{\bf s}\cdot{\bf n},
\end{equation}
\begin{equation}
\label{3.8}
\lbrack{C_1,{\bf n}}\rbrack=0,\quad\lbrack{C_2,{\bf n}}\rbrack=0.
\end{equation}
 and as a complete set of commuting operators one can select the invariants $C_1,{C}_2$ and the vector  ${\bf n}$.

In the relativistic mechanics we must   find the property of the  vector  ${\bf n}$  along the  trajectory of the particle.
In accordance with  formulas (\ref{3.5}) in the relativistic mechanics the  quantity ${\bf N}$, the field ${\bf N}(x)$ and the  invariant ${C_1}(x)$ can be expressed in the form
\begin{equation} 
\label{3.9}
{\bf N}:=\alpha{\bf n}+{\bf n}\times{\bf L}({\bf n}),\quad{C_1}=\alpha^2,
\end{equation}
\begin{equation} 
\label{3.10}
{\bf N}(x):=\alpha(x){\bf n}(x)+{\bf n}(x)\times{\bf L}(x),\quad{C_1}(x)=\alpha^2(x).
\end{equation}
Using  equations (\ref{2.11}) and (\ref{2.12}) we obtain
\begin{equation}
\label{3.11} 
\alpha^2(t_1,{\bf x}_1)=\alpha^2(t_2,{\bf x}_2),\quad{\bf n}(t_1,{\bf x}_1)={\bf n}(t_2,{\bf x}_2).
\end{equation}

Let $|{\bf n},{\lambda},{\alpha}>$ be the states with a well-defined value of the operator $C_1$ ,\,$C_2$  and the vector ${\bf n}$. Then in accordance with  Eqs.(\ref{2.12}) and (\ref{3.11}) for the transition amplitude  for a  massive relativistic particle  we have the expression in terms of the matrix elements of the unitary  irreducible representations of the Lorentz group
\begin{eqnarray}
\label{3.12}
K(x_2;x_1,{\alpha},{\lambda},n):=<{\alpha},{\lambda},{\bf n}|S(x_2-x_1)|{\bf n}^{'},{\lambda}^{'},{\alpha}^{'}>{_{{\alpha},{\lambda},{\bf n}={\alpha}^{'},{\lambda}^{'},{\bf n}^{'}}}.
\end{eqnarray}
For example for a particle with spin zero 
\begin{eqnarray}
\label{3.13}
K(x_2;x_1,n)&=&{\lefteqn({\int\frac{d{\bf p}}{p_0}\,{\xi}^{\ast(0)}({\bf p},{\alpha},{\bf n})\,S(x_2-x_1){\xi}{^0}({\bf p},{\alpha}^{'},{\bf n^{'}})})}{_{{\alpha},{\lambda},{\bf n}={\alpha}^{'},{\lambda}^{'},{\bf n}^{'}}}\nonumber\\&&=\frac{1}{(2\pi)^{3}}\int\frac{d{\bf p}}{p_0}\frac{\exp-i[(x_2-x_1)p]}{[(pn)/m]^2}.
\end{eqnarray}
This transition amplitude  contain  the vector  of the  light-cone ${n}$. 
Applying the operator $[i(n_0{\partial}_{{t}}+{\bf n}{\nabla}_{\bf x})/m]^2$  we have the relation to the Feynman propagator ${\triangle}^{+}(x)$ of the free Klein-Gordon  equation 
\begin{equation}
\label{3.14}
{\triangle}^{+}(x)=\frac{-i}{2}[i(n_0{\partial}_{{t}}+{\bf n}{\nabla}_{\bf x})/m]^2K(x,n),
\end{equation}
where
\begin{equation}
\label{3.15}
{\triangle}^{+}(x)=\frac{-i}{(2\pi)^{3}}\int\frac{d{\bf p}}{2p_0}{\exp-i[px]}.
\end{equation}
In the nonrelativistic limit in the momentum representation ${\bf N}\rightarrow{im{\nabla}_{\bf p}:={\bf q}}$ and
\begin{equation}
\label{3.16}
{\xi}^{(0)}({\bf p},{\alpha},{k}){\;}\rightarrow{\;}\Psi({\bf p},\alpha{\bf n}):=\frac{1}{(2\pi)^{3/2}}\exp[-i(\alpha{\bf n})\cdot{\bf p}/m].
\end{equation}
 The functions $\Psi({\bf p},\alpha{\bf n})$ are the eigenfunctions of the operators ${\bf q}$ and ${\bf q}^2$. 
In \cite{Frick3})  was remarked that  in the expansion  of the Galileo group
\begin{equation}
\label{3.17}
\Psi(\alpha{\bf n})={\frac{1}{{(2\pi)}^{3/2}}}\int{d{\bf p}}\Psi({\bf p})\exp(i{\alpha\bf n}\cdot{\bf p}/m),   
\end{equation}
where $\Psi({\bf p})$ and $\Psi({\alpha}{\bf n})$
are the states of the particle in ${\bf p}$ and in the ${\alpha},{\bf n}$ representation the kernel $\exp(i{\alpha\bf n}\cdot{\bf p}/m)$
 can be replaced by plane waves  $\exp(i{\bf x}\cdot{\bf p})$ and in  such a form the
${\bf x}$ representation in the nonrelativistic limit can be constructed. It is well known that   the transition amplitude  may be written in this case as $<{{\bf x}_{2}|S(t_{2}-t_{1})|{\bf x_1}}>$.
In the relativistic  region  this method cannot be used. The  functions   which  realize the  unitary irreducible space-time independent representations of the Lorentz group and  of the Galileo group   have different forms.

In the framework of the  generalized Heisenberg/Schr\"odinger picture   for describing  the motion of a  particle  in  the nonrelativistic limit   we can use the same method as in the  relativistic case.  The operators
\begin{equation}
\label{3.18}
{\bf q}(t,{\bf x})={\bf q}+t{\bf p}-{\bf x}{m},\quad{\bf q}(t,{\bf x})={\bf q}(t_0,{\bf x}_0)+(t-t_0){\bf p}-({\bf x}-{\bf x}_0){m},
\end{equation}
just as ${\bf N}(x)$ can be viewed as field operators with the property
\begin{equation}
\label{3.19}
\frac{\partial{q_i(t,{\bf x})/m}}{\partial{t}}=P_i/m,\quad\frac{\partial{q_i(t,{\bf x})/m}}{\partial{x_j}}=-{\delta}_{ij}.
\end{equation}
In order to find the transition amplitude we  must calculate the matrix element
$(H={\bf p}^2/2m)$
\begin{equation}
\label{3.20}
K(x_{2};x_{1},{\alpha}{\bf n}):=<{\alpha}{\bf n}|S(t_{2}-t_{1},{\bf x}_{2}-{\bf x}_{1})|{\bf n}^{'}{\alpha}^{'}>{_{{\alpha}{\bf n}={\alpha}^{'}{\bf n}^{'}}}
\end{equation}
which  correspond to the  transition $S(t_{2}-t_{1},{\bf x}_{2}-{\bf x}_{1})$ of the particle between states with equal eigenvalues of the operator ${\bf q}$ and ${\bf q}^2$ . Using  the functions $\Psi({\bf p},\alpha{\bf n})$  we obtain expression 
\begin{eqnarray}
\label{3.21}
K(x_{2};x_{1},\alpha{\bf n})&=&{\lefteqn({\int{d{\bf p}}\Psi^{\ast}({\bf p},\alpha{\bf n})S(t_{2}-t_{1},{\bf x}_2-{\bf x}_1)\Psi({\bf p},{\alpha^{'}}{\bf n}^{'})})}{_{{\alpha}{\bf n}={\alpha}^{'}{\bf n}^{'}}}\nonumber\\&&={{\lbrack\frac{m}{2\pi\imath(t-t_0)}\rbrack}}^{3/2}\exp\frac{\imath{m}({\bf x}_{2}-{\bf x}_{1})^2}{2(t_{2}-t_{1})}.
\end{eqnarray}
which agree with the transition amplitude in  quantum  mechanics.
\section{CONCLUSION}
We have shown that  the propagation of a relativistic particle may be  described in terms of the   transformation $S(x)=\exp[-i(tH-{\bf x}\cdot{\bf P})]$  and the matrix elements of the unitary  irreducible representations of the Lorentz group.  We have considered the operators of the Lorentz algebra in the generalized Heisenberg picture as field operators and found that the analogue of the space-time independent operators  in the relativistic mechanics must be separated from the integrals  of  motion. In this case the  conversion to the quantum version can take place. The transition amplitude for a particle with spin zero  contain  a vector  of the  light-cone  which appears  in the  expansions  of the Lorentz group. Finally we have shown that in  the nonrelativistic limit the transition amplitude may be also expressed in terms of the transformation $S(x)$.
\subsubsection*{ACKNOWLEDGMENTS}

This work was supported by the Deutsche Forschungsgemeinschaft (No. FR 1560/1-1).

\end{document}